# Optimizing performance for on-chip SBS-based isolator

CHOON KONG LAI,[1,2,*] MORITZ MERKLEIN,[1,2] ALVARO CASAS-BEDOYA,[1,2]
YANG LIU,[3] STEPHEN J. MADDEN,[4] CHRISTOPHER G. POULTON,[5]
MICHAEL J. STEEL[6], BENJAMIN J. EGGLETON[1,2]

[1]*The University of Sydney Nano Institute (Sydney Nano), The University of Sydney, NSW 2006, Australia*
[2]*Institute of Photonics and Optical Science (IPOS), School of Physics, The University of Sydney, NSW 2006, Australia*
[3]*Institute of Physics, Swiss Federal Institute of Technology Lausanne (EPFL), CH-1015, Lausanne Switzerland*
[4]*Deparment of Quantum Science and Technology, Research School of Physics, Australian National University, Acton ACT 2601, Australia*
[5]*Institute School of Mathematical and Physical Sciences, University of Technology Sydney (UTS), NSW 2007, Australia*
[6]*MQ Photonics Research Centre, School of Mathematical and Physical Sciences, Macquarie University, NSW 2109, Australia*

*\*choon.lai@sydney.edu.au*

**Abstract:** Non-reciprocal optical components such as isolators and circulators are crucial for preventing catastrophic back-reflection and controlling optical crosstalk in photonic systems. While non-reciprocal devices based on Brillouin intermodal transitions have been experimentally demonstrated in chip-scale platforms, harnessing such interactions has required a suspended waveguide structure, which is challenging to fabricate and is potentially less robust than a non-suspended structure, thereby limiting the design flexibility. In this paper, we numerically investigate the performance of a Brillouin-based isolation scheme in which a dual-pump-driven optoacoustic interaction is used to excite confined acoustic waves in a traditional ridge waveguide. We find that acoustic confinement, and therefore the amount of Brillouin-driven mode conversion, can be enhanced by selecting an appropriate optical mode pair and waveguide geometry of two arsenic based chalcogenide platforms. Further, we optimize the isolator design in its entirety, including the input couplers, mode filters, the Brillouin-active waveguide as well as the device fabrication tolerances. We predict such a device can achieve 30 dB isolation over a 38 nm bandwidth when 500 mW pump power is used; in the presence of a $\pm$ 10 nm fabrication-induced width error, such isolation can be maintained over a 5-10 nm bandwidth.

## 1. Introduction

Miniaturizing functionality that allows unidirectional transmission of optical signals is one of the key priorities in photonics research. This requires breaking Lorentz reciprocity [1] which has traditionally been achieved by Faraday rotation in magneto-optic garnets [2]. While on-chip integration is challenging because of apparent incompatibilities such as high lattice and thermal mismatch with common semiconductor substrates, there has been consistent progress in pursuing integrated magnetic isolators by means of hybrid bonding [3–7] or direct-deposition method [8,9], with impressive results ― 3 dB insertion loss (IL) and 40 dB isolation ratio (IR) ― in the latter case [9]. However, such an isolator relies on the nonreciprocal phase shift between the forward and backward propagation modes in a resonator, and therefore the operational bandwidth is very limited.

On the other hand, magnetic-free isolators can be realized by exploiting second-order [10] or third-order (Kerr) optical nonlinearities [11–18]. These nonlinear isolators rely on the injection of substantial power in the forward direction to prevent backward light propagation; as a result, they can only operate in a specific signal power range and are incapable of blocking a weak backward signal in the case of intense forward power [19]. Recent research has also focused on inducing spatiotemporal modulation of light in a non-magnetic optical waveguide [19–37]. This class of non-reciprocity relies on intermodal photonic transitions driven by electro-optic or acousto-optic effects. Excellent performance of acousto-optic non-reciprocity has been reported in aluminium nitride on silicon rib waveguides [34], with a < 0.6 dB IL, ~16 dB IR and > 100 GHz operational bandwidth.

An alternative route to acousto-optic non-reciprocity is to use stimulated Brillouin scattering (SBS) — a phenomenon in which an optically driven acoustic wave promotes an optical modal transition with a typical frequency shift in the GHz range. Such transitions can be either intramodal or intermodal, and occur in the same or opposite propagation direction of the optical pump depending on the phase matching condition of both optical modes involved. The four types of SBS are often referred to as backward SBS [38–44], backward intermodal SBS (BIBS) [45], forward SBS (FBS) [46] and forward intermodal SBS (FIBS) [47]. Among them, FIBS has been proposed in the earlier Brillouin isolation schemes in optical fibers [21] and planar waveguides [23] because of its large operational bandwidth arising from the almost parallel dispersion profile between the co-propagating fundamental and higher order optical spatial modes, such that phase matching between them can be maintained over a large wavelength range. Later, isolators based on non-local interband Brillouin scattering (NIBS) have been experimentally demonstrated in a suspended silicon structure [33]. Such NIBS process is generated by a pair of asymmetrical rib waveguides placed in close proximity, where FIBS is first excited by pumping one of the waveguides, and the mediating shear waves — known as Lamb waves [48] — travel to the adjacent waveguide through the silicon slab, allowing a similar optical intermodal transition to occur at the adjacent waveguide which carries the optical signal.

Suspended structures however are sophisticated waveguide structures requiring complex fabrication processes; furthermore, the integration of these waveguides in future multi-layer structures may prove challenging. In addition, silicon suffers from nonlinear losses and free-carrier absorption, which in turn limits the net Brillouin amplification [48,49]. It would be highly advantageous therefore to have an entirely embedded structure in which these non-reciprocal Brillouin effects may be harnessed in a material that lacks nonlinear loss. However, such an embedded structure itself presents considerable design challenges.

Foremost, the performance of the isolator is highly dependent on the presence of shear waves in the waveguide. The effect of shear components was assumed to be negligible in earlier studies [23] but will affect both the acoustic confinement as well as the overall Brillouin gain. Rib waveguides, which are highly successful for backwards SBS in embedded structures, also support Lamb waves that can transport energy away from the core and are unsuitable for acoustic guidance in the forward direction. On the other hand, embedded waveguides also support a wide range of optical and acoustic modes, all of which will possess different coupling, losses, and Brillouin gain. In addition, a functional isolator must also include a mode coupler that can be integrated into the chip structure for exciting the desired optical spatial modes in the Brillouin active waveguide. The combination of waveguide and coupler design, together with fabrication constraints, together with the necessary conditions for an isolator to be useful, presents a challenging and complex design problem.

This article provides a detailed design strategy for realizing an isolation scheme based on forward intermodal Brillouin scattering (FIBS) in a non-suspended waveguide. We carry out a rigorous study of the complete device using established photonics simulation tools to examine the acoustic confinement and optimize the performance of each on-chip component involved. By performing full electromagnetic simulations of the couplers, and combining this with accurate Coupled-Mode analysis of the SBS process in the Brillouin-active waveguide, we demonstrate that the acoustic guidance can be enhanced by selecting an appropriate combination of rectangular waveguide dimensions and pairs of optical spatial modes. We use our design strategy to optimize an isolator structure in an embedded chalcogenide ridge waveguide; we predict such a device can achieve 30 dB isolation over a 38 nm bandwidth for moderate pump powers (~500 mW). Our strategy of mode selection also allows us control over the bandwidth of the device: we show that our design maintains isolation over a 5 nm (10 nm) bandwidth at -10 nm (+10 nm) fabrication-induced width error with a reasonable input optical power of 500 mW. The choice of a +/- 10 nm width deviation in our calculation is within the achievable precision of advanced fabrication tools such as electron beam lithography (EBL), dry-etching system, scanning electron microscope (SEM), etc. and thus it is a conservative guideline to follow in the future fabrication work. These structures therefore offer a viable alternative route to non-reciprocity, based on fully-embedded waveguides.

The design concept of the whole device and its working principle is first described in the following section. Section 3 focuses on interrogating the key quantities (Brillouin shift, acoustic phase velocity $v_p$ and gain coefficient) for different spatial optical mode pairs at different waveguide dimensions of two different chalcogenide platforms ($As_2S_3$ and $As_2Se_3$). In Section 4 we show how the isolation bandwidth can be optimized by tailoring the waveguide dispersion. Further, we compute overall device performance after incorporating a broadband and dimensional tolerant mode coupler. Finally, a method for improving the device isolation is proposed in Section 5.

## 2. Working Principle

The Brillouin-based non-reciprocal effect in this work relies on the FIBS process [50], which can be summarized by the processes in the schematic dispersion diagram in Fig. 1. The two pump channels $\omega_1$ and $\omega_2$ described in the figure may occupy either different spatial modes or polarization states [47]. Then the acoustic modes mediating the transition between the two optical mode of interest can be of different symmetry and character (quasi-flexural, quasi-torsional and quasi-longitudinal modes) [51]. This work focuses on different optical spatial modes with the same polarization state because, as outlined in Section 3, inter-polarization Brillouin coupling is incapable of generating sufficiently large acoustic wavenumbers to fulfill the acoustic wave guiding requirements.

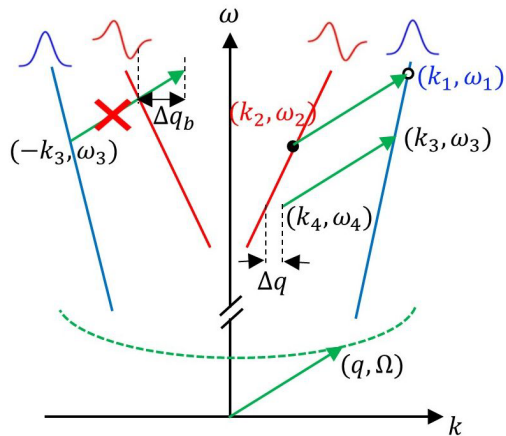

Fig. 1. Dispersion diagram describing the Brillouin-based non-reciprocal effect.

In the proposed isolation scheme, two co-propagating optical waves at angular frequencies $\omega_1$ and $\omega_2$ induce a coherent acoustic oscillation at frequency $\Omega$ and wavenumber $q$ through electrostriction and radiation pressure. Pump $\omega_1$ scatters off this dynamic grating and experiences a Brillouin shift to $\omega_2$. The blue and red curves in Fig. 1 indicate slower and faster propagating optical modes. By phase-matching arguments, the frequency and wavenumbers of the three waves are related by $\Omega = \omega_1 - \omega_2$ and $q = k_1 - k_2$. For a specific waveguide structure, the Brillouin shift $\Omega$ at a particular acoustic wavenumber is determined by the acoustic dispersion curve (the green dashed curve). Meanwhile, a third co-propagating optical wave $\omega_3$ with weak amplitude and small frequency separation from $\omega_1$ can be scattered by the same acoustic wave and experience the same Brillouin shift to $\omega_4 = \omega_3 - \Omega$ provided the phase mismatch, $\Delta q = k(\omega_3) - k(\omega_4) - q$ is negligible. On the other hand, the corresponding scattering of a counter-propagating optical wave $\omega_3$ is forbidden owing to a large phase mismatch ($\Delta q_b \gg \Delta q$).

The proposed design of the integrated isolator is depicted in Fig. 2(a). The whole device comprises a pair of reciprocal mode couplers MC1 and MC2, a multimode waveguide (the non-reciprocal mode coupler (NRMC)) and several mode-size converters linking different optical components. Two continuous pump waves of frequency $\omega_1$ and $\omega_2$ are launched into Port 1 and Port 3 respectively. Ideally, light entering all ports propagates in the fundamental optical spatial mode in the waveguide. After propagating through the first mode multiplexer MC1, the mode nature of pump $\omega_1$ remains unchanged while pump $\omega_2$ undergoes mode conversion to a higher-order mode via linear coupling. The intensity beat note of pump $\omega_1$ and $\omega_2$ induces an acoustic wave that enables coherent forward scattering, whereby a co-propagating signal wave $\omega_3$ injected from Port 1 is converted to a higher-order optical mode at $\omega_4$, which is then demultiplexed at the second mode coupler MC2 and output at Port 4. The output power from Port 2 is negligible if the mode conversion efficiency from both the FIBS process and mode demultiplexing at MC2 approaches 100%. On the other hand a signal wave at $\omega_3$ input from Port 2 will be transmitted to Port 1 without experiencing any FIBS-induced mode conversion.

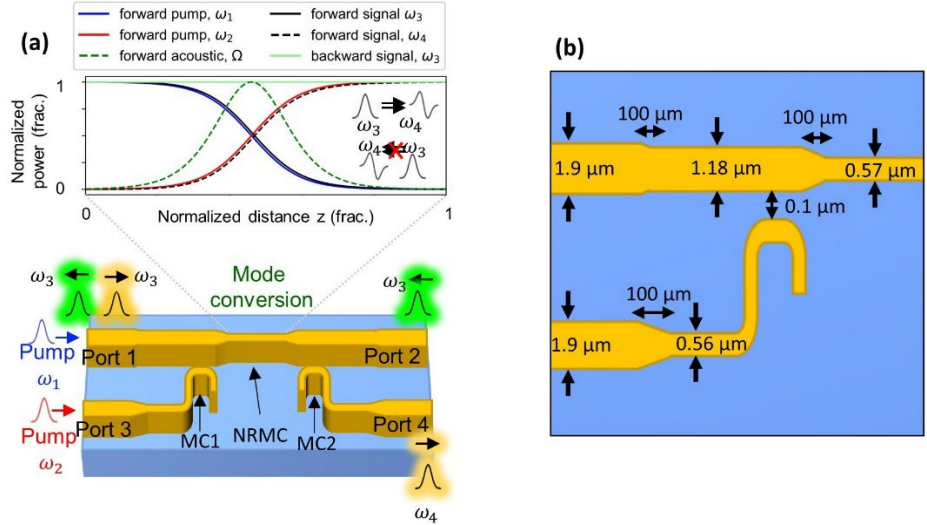

Fig. 2. (a) Schematic of the Brillouin-based optical isolator. The plot in the figure illustrates the normalized power of all optical and acoustic waves travelling at each position of the non-reciprocal mode coupler NRMC. (b) Top view showing the waveguide width variation in the first half of the proposed isolator device.

It has been shown previously [23] that a better nonreciprocal mode conversion efficiency can be achieved by increasing the initial power ratio of the forward pump $\omega_1$ to that of the pump $\omega_2$ such that the power of the forward acoustic wave peaks at the middle of the NRMC [23]. The plot in Fig. 2(a) illustrates a typical power variation of the pump and signal waves along the NRMC with an initial pump power ratio of ~0.99. The waveguide width variation in the proposed design is clearly illustrated in Fig. 2(b). Port 1, 2, 3 and 4 have a horizontal width $w$ (~1.9 μm) that is mode-match to a lens-tipped fibre with Gaussian spot size of ~2 μm and minimizes sidewall scattering loss that is particularly relevant when a long integrated circuit is required for future applications. Therefore, two different spot size converters (SSC) are used to narrow down the horizontal width $w$ of Port 1 and Port 3 into the required dimensions of the parallel waveguides in MC1, and a horizontal taper is used to further shrink the waveguide into the width required by the NRMC. Likewise, another half of the device requires similar horizontal tapers and SSCs to expand the waveguide into the width of Port 3 and 4.

## 3. Optical and acoustic confinement in a non-suspended chalcogenide waveguide

The primary challenge of the proposed Brillouin isolation scheme is to achieve strong optical and acoustic guidance for co-propagating waves in a non-suspended waveguide. For both optical and acoustic waves, this can be realized by ensuring that the mode has a lower phase velocity in the waveguide core than that of all propagating modes in the cladding. Various chalcogenide platforms such as GeSbS [52] and $As_2S_3$ [38–44] can achieve this in backward SBS process as they exhibit a lower phase speed of both light and sound compared with silica, which can be used as a cladding material. However, as illustrated in Fig.1, the phase matching condition in the forward SBS process results in a much smaller $q$ (longer elastic wavelengths) as compared with the backward counterparts. Such long elastic wavelengths involved in the FIBS process pose a major challenge: these wavelengths are very close to or even beyond the elastic mode cut-off regime, making acoustic confinement more difficult.

Our numerical modelling of acoustic modes suggests that standard shallow-etched $As_2S_3$ rib waveguides do not possess appropriate acoustic guided modes for FIBS. However, such guided modes can be found with a further increase in the etch depth or using a fully etched (strip) waveguides. Such improvement in the acoustic guidance is achieved by increasing the acoustic wavenumber $q$, which is determined by the effective index difference between the two optical spatial modes of interest if the optical wavelength is fixed. This also suggests that, apart from the waveguide dimensions and choices of optical mode pairs, selecting a higher refractive index material with similar acoustic properties could further improve the acoustic confinement.

Table 1. Optical and elastic properties of $As_2S_3$, $As_2Se_3$ and $SiO_2$ used in the NumBAT simulation where $n$ and $\rho$ are refractive index at 1.55 μm wavelength and material density respectively. $c_{ij}$, $p_{ij}$, and $\eta_{ij}$ are fourth rank stiffness, photoelastic and phonon viscosity tensors in units of mPa.s expressed in the Voigt compact notation. [53]

| Material | n @ 1.55 μm | $\rho$ (kgm$^{-3}$) | $c_{11}$ (GPa) | $c_{12}$ (GPa) | $c_{44}$ (GPa) | $p_{11}$ | $p_{12}$ | $p_{44}$ | $\eta_{11}$ | $\eta_{12}$ | $\eta_{44}$ |
|---|---|---|---|---|---|---|---|---|---|---|---|
| $SiO_2$ [53–56] | 1.45 | 2200 | 78.6 | 16.1 | 31.2 | 0.12 | 0.27 | -0.075 | 1.6† | 1.29† | 0.16† |
| $As_2S_3$ [56–58] | 2.44 | 3150 | 19.75 | 8.7 | 5.52 | 0.25 | 0.23 | 0.01 | 1.8† | 1.45† | 0.18† |
| $As_2Se_3$ [50,52,55,56] | 2.84 | 4635 | 23.5 | 9.5 | 7.0 | 0.31 | 0.27 | 0.02 | 0.78† | 0.63† | 0.08† |

† is the theoretical estimate using the Smith et al. approach [56]

To investigate the impact of the geometric parameters on the acoustic confinement, we solve the elastic and optical mode problems and find the Brillouin gain coefficients for waveguides of several different dimensions of silica-clad arsenic sulfide ($As_2S_3$) and a higher-index arsenic selenide ($As_2Se_3$) multimode waveguides using the open-source Numerical Brillouin analysis tool (NumBAT) [61,62], which has been validated against a number of reported experimental results [63–66]. Arsenic trisulfide ($As_2S_3$) is a mature on-chip SBS platform that has been substantially studied for a decade [38–44]. While on-chip SBS has not yet been demonstrated in the selenide platform ($As_2Se_3$), a high SBS gain has been observed in selenide fibres [54]. Together with its high refractive index, excellent acoustic guidance is anticipated in the same waveguide dimension.

The relevant physical properties of $As_2S_3$ and $As_2Se_3$ are shown in Table 1. From the stiffness tensor components $c_{ij}$, the longitudinal ($v_l$) and shear speeds of sound ($v_s$) in each material can be computed using the relations

$$v_l = \sqrt{\frac{c_{11}}{\rho}}, \quad v_s = \sqrt{\frac{c_{44}}{\rho}} \tag{1}$$

A parametric sweep of waveguide dimensions was executed to study the strength of acoustic guidance. For silica-clad sulfide ($As_2S_3$) waveguides, a width $w$ scan from 600 nm to 3000 nm was performed at three different vertical thicknesses 500 nm, 700 nm, and 1000 nm. To investigate the effect of refractive index on the acoustic confinement, we performed a similar width scan on a higher index selenide ($As_2Se_3$) waveguide with 700 nm vertical thickness. Furthermore, while it is likely more feasible to induce FIBS with the two lowest order modes $TE_{11}$ and $TE_{21}$, the elastic mode confinement arising from other higher-order spatial modes listed in Fig. 3 is also computed to identify the best possible confinement and gain for each geometry.

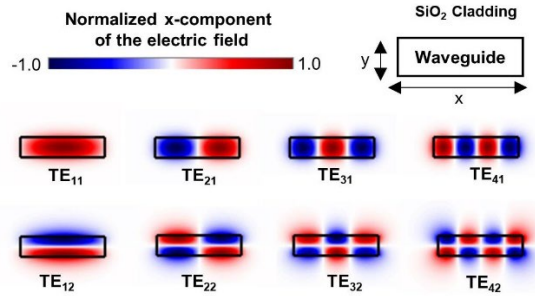

Fig. 3. Normalized electric field distribution of different quasi-transverse electric (TE) spatial modes in a rectangular waveguide. They are labelled in a conventional way as $TE_{pq}$ where p and q indicate the number of horizontal and vertical lobes respectively.

In general, the degree of optical confinement can be described by the normalized propagation constant $\beta$, which is given by

$$\beta = \frac{n_{\text{eff}} - n_{\text{cl}}}{n_{\text{co}} - n_{\text{cl}}} = \frac{v_{\text{co}}}{v_{\text{eff}}}\left(\frac{v_{\text{cl}} - v_{\text{eff}}}{v_{\text{cl}} - v_{\text{co}}}\right) \tag{2}$$

where it can be expressed in terms of refractive indices or phase velocity. $n_{\text{eff}}$, $n_{\text{co}}$ and $n_{\text{cl}}$ are effective mode index, core and cladding refractive index. The corresponding phase velocities $v_{\text{eff}}$, $v_{\text{co}}$ and $v_{\text{cl}}$ can be determined by the refractive index relation $n = c/v$ where $c$ is the speed of light in vacuum. $\beta$ is a good indicator of the optical confinement and has a value ranging from 0 to 1. For instance, $\beta = 1$ means the optical field completely confined in the waveguide core while $\beta = 0$ represents zero confinement.

Likewise, a normalized propagation constant can be used to describe the acoustic confinement. The concept of using normalized propagation constant as a confinement metric for both optical and acoustic modes is depicted in Figs. 4(a-b). In comparison, the elastic wave propagation is rather complicated due to the existence of both transverse (shear) and longitudinal (compressive) mechanical field nature that give rise to different sound velocity in bulk materials. Therefore, instead of one light line for each material in the optical dispersion (Fig. 4 (a)), two sound lines corresponding to the shear and longitudinal wave nature are drawn on the acoustic dispersion plot (see Fig. 4(b)). For an embedded rectangular waveguide, a confined acoustic mode requires an effective phase velocity $v_\mathrm{p}$ smaller than the cladding shear speed $v_\mathrm{s,cl}$ while any numerical solutions with $v_\mathrm{p} > v_\mathrm{s,cl}$ are considered as a free mode [67], in which energy is no longer being confined in the core; the confinement increases as $v_\mathrm{p}$ decreases from the cladding shear line. To facilitate the following discussion, we define the normalized acoustic propagation constant $\beta_a$ as

$$\beta_\mathrm{a} = \frac{v_\mathrm{l,co}}{v_\mathrm{p}} \left( \frac{v_\mathrm{s,cl} - v_\mathrm{p}}{v_\mathrm{s,cl} - v_\mathrm{l,co}} \right) \tag{3}$$

where $v_\mathrm{l,co}$ represents the core longitudinal speed. It is noteworthy that while the acoustic guidance is analogous to that of the optical counterpart, $\beta_\mathrm{a} > 1$ is possible due to the existence of shear waves that propagate at a slower speed than the longitudinal (compressive) waves in the core ($v_\mathrm{s,co} < v_\mathrm{l,co}$), resulting in $v_\mathrm{p} < v_\mathrm{l,co}$.

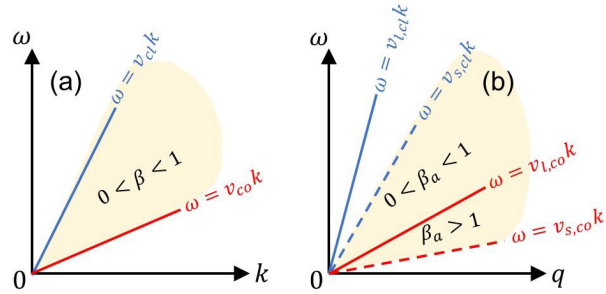

Fig. 4. (a) Optical and (b) acoustic dispersion plot showing the light and sound lines for the core and cladding material. The yellow region indicates the area where a confined mode can be found.

In Figs. 5(a-d), we plot the normalized propagation constant $\beta$ versus $w$ for different waveguide structures and optical spatial modes at an optical pump wavelength of 1550 nm. In Figs. 5(e-h) we investigate the behavior of different acoustic modes as a function of waveguide width $w$. We found that all numerical solutions within the parameter sweep range in Fig. 5 have $\beta_\mathrm{a} < 1$. At a specific dimension and optical mode pair, there exists more than one solution that lies in the confinement window, from which we only select the mode that records the highest Brillouin gain coefficient $g_0$ to plot in Figs. 5(e-h). The strength of the Brillouin coupling $g_0$ for each of these modes is computed through standard expressions based on optoacoustic mode overlap integrals [61,62].

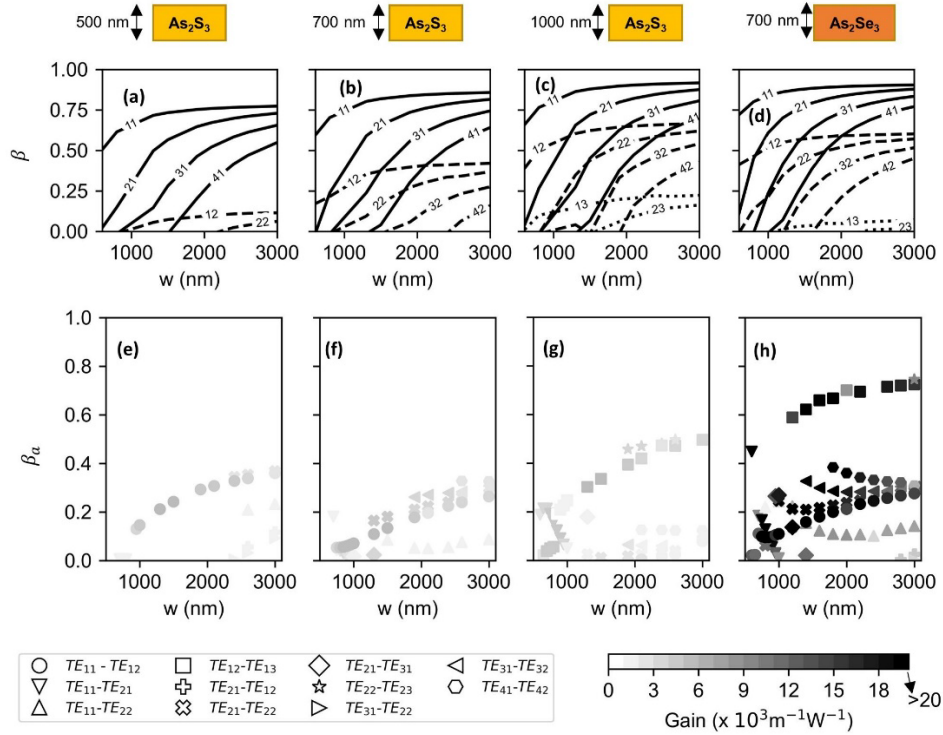

Fig. 5. Optical and acoustic confinement of $As_2S_3$ and $As_2Se_3$ waveguides: (a-d) Optical dispersion diagram showing $\beta$ variation with respect to the width $w$ of $As_2S_3$ waveguides at a vertical thickness of (a) 500 nm, (b) 700 nm, and (c) 1000 nm, and (d) $As_2Se_3$ waveguides at 700 nm thickness for different quasi-transverse electric spatial modes $TE_{pq}$ where the mode label $pq$ is indicated by the inline label. Solid, dashed and dotted lines are used to indicate $q = 1$, $q = 2$, and $q = 3$ respectively. (e-h) Plots of $\beta_a$ of the guided acoustic modes versus $w$ corresponding to the waveguide structures of (a-d) respectively. The shape of the scatter points represents different optical mode pairs in the bottom legend while the sequential color scales indicate the gain coefficient in units of $m^{-1}W^{-1}$ according to the colorbar.

From the optical dispersion diagram in Figs. 5(a-d), one can see that the confinement for each $TE_{pq}$ modes improves with the waveguide size. Further, the number of guided $TE_{pq}$ modes — with normalized propagation constant above the so-called "cladding light line" ($\beta > 0$) — increases when $w$ expands from 600 nm to 3000 nm, leading to a higher number of mode combination choices at a larger $w$. However, the trend of acoustic confinement in Figs. 5(e-h) can be quite different from that of the optical counterpart. This is because, apart from waveguide dimension, $\beta_a$ is also dependent on the normalized propagation constant difference between two optical mode of interest $\Delta\beta$, due to the fact that $q = k_1 - k_2$ as described in Fig. 1. For instance, for 500 nm thick $As_2S_3$ waveguides (Fig. 5(e)), the abrupt increase of $\beta_a$ corresponding to $TE_{11}$-$TE_{12}$ pair (●) at $w < 1200$ nm is likely due to the high $\Delta\beta$ between the $TE_{11}$ and $TE_{12}$ modes in Fig. 5(a). For $w > 1200$ nm, since there is no further increase in $\Delta\beta$, hence the increase of $\beta_a$ in this width regime is solely caused by the increase of waveguide size. This relation between $\Delta\beta$ and $\beta_a$ is also true for all other mode combinations and it explains why there is actually a decrease of $\beta_a$ for $TE_{11}$-$TE_{21}$ (▼), $TE_{11}$-$TE_{22}$ (▲), and $TE_{41}$-$TE_{42}$ (⬤) pairs for a given $w$ range in Fig. 5(h). While it can be deduced from plots (e-h) that the highest possible $\beta_a$ among the available mode pairs increase with $w$, the impact of vertical thickness on the maximum $\beta_a$ is not obvious from plots (e-g).

From Figs. 5(d,h) we see that switching to a higher index selenide platform (As$_2$Se$_3$) can significantly improve the acoustic confinement, from $\beta_a \approx 0.3$ to $\beta_a \approx 0.7$ (compare plot (f) with (h)). This is aligned with the previous observation that higher refractive index results in a higher $q$ or $\Delta\beta$, and so should increase the acoustic confinement. Together with a larger photo-elastic coefficient, the Brillouin gain coefficient in the selenide (As$_2$Se$_3$) waveguide is approximately a factor of three over the sulfide (As$_2$S$_3$) platform. Despite the fact that such high theoretical gains can only be achieved if the processed waveguides are free from sidewall roughness-induced losses, the material's intrinsic optical and acoustic absorption and other detrimental factors, the gain improvement by a factor of 3 from the sulfide to selenide platforms is still promising provided their optical and acoustic losses are comparable.

By comparing Figs. 5(e-h), it can be seen that the acoustic mode from the TE$_{12}$-TE$_{13}$ transition (■) exhibits excellent acoustic confinement in both material systems, with $\beta_a \approx 0.7$ in the selenide platform (see Figs. 6(i-l)). However, it is advantageous to use the lowest order mode pairs possible because the higher order counterparts often experience a greater fabrication-induced sidewall scattering loss and crosstalk, and can be difficult to couple into. The acoustic confinement for different dimensions for both lowest order pairs TE$_{11}$-TE$_{21}$ (▼) and TE$_{11}$-TE$_{12}$ (●) are shown in Fig. 5. We see that for the TE$_{11}$-TE$_{21}$ pair (▼), higher $\beta_a$ is observed at a higher aspect ratio waveguide (small $w$ and large $t$). On the other hand, it is preferable to have a lower aspect ratio waveguide for the TE$_{11}$-TE$_{12}$ pair (●).

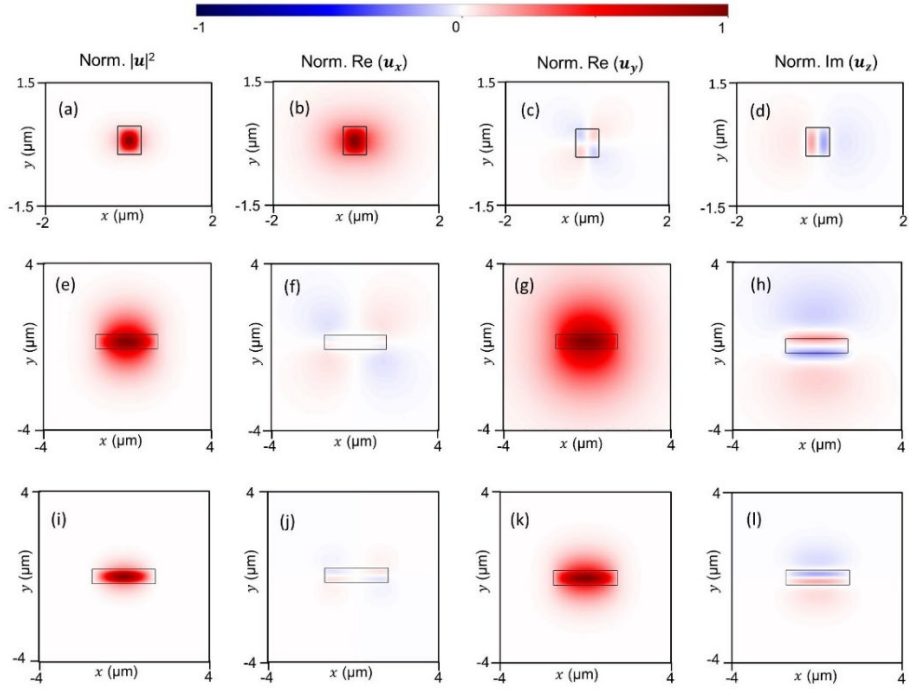

Fig. 6. The displacement field, $\boldsymbol{u}$ profile in the $\boldsymbol{x}$-$\boldsymbol{y}$ plane of 700 nm thick As$_2$Se$_3$ waveguide with (a-d) TE$_{11}$-TE$_{21}$ pair at $\boldsymbol{w} = 570$ nm, (e-h) TE$_{11}$-TE$_{12}$ pair at $\boldsymbol{w} = 3000$ nm, and (i-l) TE$_{12}$-TE$_{13}$ pair at $\boldsymbol{w} = 3000$ nm. The first column (a,e,i) shows the square of the magnitude of the complex amplitude $|\boldsymbol{u}|^2$, followed by the real part of its $\boldsymbol{x}$-component (b,f,j), the real part of its $\boldsymbol{y}$-component (c,g,k), and the imaginary part of its $\boldsymbol{z}$-component (d,h,l).

While all acoustic modes presented in Fig. 5 have $\beta_a < 1$, the displacement field $\boldsymbol{u}$ profile in Fig. 6 illustrates that they have strong shear (transverse) field components. The polarization of the transverse displacement field is dependent on the types of optical mode transition. In general, TE$_{pq}$-TE$_{(p+1)q}$ transition can induce a forward propagating acoustic wave with a strong lateral component of $\boldsymbol{u}_x$ (see Figs. 6(a-d)) whereas the acoustic wave from $TE_{pq}$-$TE_{p(q+1)}$ transition has a strong $\boldsymbol{u}_y$ component (see Figs. 6(e-h) and 6(i-l)).

Altering the waveguide dimension can significantly change the confinement factor of such $x$- and $y$-polarized acoustic modes. Focusing on Figs. 5(g-h), the $x$-polarized acoustic mode due to the $TE_{11}$-$TE_{21}$ transition loses its confinement when the horizontal width increases. Meanwhile, to suppress the $y$-polarized acoustic mode due to the $TE_{11}$-$TE_{12}$ transition, one can increase the vertical thickness (see Figs. 5(e-g)). It is also worth mentioning that while the confinement of both acoustic modes can be improved by reducing the width and the thickness, further dimensional shrinkage ($w < 500$ nm and $t < 500$ nm) can lead to a converse effect as the optical modes can no longer be confined in such a small waveguide core.

## 4. Overall performance of the proposed isolator

In previous section, we have identified the optical mode pairs and waveguide dimensions of two binary chalcogenide platforms in which confined acoustic modes can be excited to yield high Brillouin gain. Next, we selectively study the narrow selenide ($As_2Se_3$) waveguide structures ($w < 600$ nm) based on the $TE_{11}$-$TE_{21}$ mode pair. This pair has been chosen because the resulting combination exhibits appreciable acoustic confinement ($\beta_a \approx 0.45$) and an excellent gain coefficient ($g_0 \approx 2.4 \times 10^4$ m$^{-1}$W$^{-1}$). We compute the overall isolator performance — isolation bandwidth — of such narrow selenide waveguide by solving the coupled mode equations and also taking into account the mode coupler device performance.

### 4.1 Solving coupled mode equations

An isolator should operate over as broad a bandwidth as is feasible. For Brillouin-type isolators as considered here, detuning of the signals from the pumps gradually increases the phase mismatch between the acoustic and optical fields $\Delta q$, resulting in reduced mode conversion when the phase mismatch becomes too large. Here we investigate this effect quantitatively, with the aim of determining the bandwidth of our entire device.

For FIBS mode conversion, the change in field amplitudes with respect to the propagation distance, $z$ of the two forward travelling pump waves ($\omega_1$ and $\omega_2$) and the two signal waves ($\omega_3$ and $\omega_4$) at a particular $\Delta q$ are described by the equations [23,61]

$$\frac{\partial a_1}{\partial z} + \frac{\alpha_1}{2} a_1 = -\frac{g_0}{2} |a_2|^2 a_1, \qquad (4)$$

$$\frac{\partial a_2}{\partial z} + \frac{\alpha_2}{2} a_2 = \frac{g_0}{2} |a_1|^2 a_2, \qquad (5)$$

$$\frac{\partial a_3}{\partial z} + \frac{\alpha_3}{2} a_3 = -\frac{g_0}{2} a_1^* a_2 e^{i\Delta qz} a_4, \qquad (6)$$

$$\frac{\partial a_4}{\partial z} + \frac{\alpha_4}{2} a_4 = \frac{g_0}{2} a_1^* a_2 e^{-i\Delta qz} a_3, \qquad (7)$$

where $|a_i(z)|^2$ represents the physical power $P_i(z)$ in Watts carried in the optical fields, $\alpha_i$ is the optical propagation loss in m$^{-1}$ and $g_0$ is the Brillouin gain coefficient in m$^{-1}$W$^{-1}$. A full derivation is provided in sections 3 and 4 of Supplement 1. Here Eqs. (4-7) describe the power flow from $P_1$ to $P_2$ and $P_3$ to $P_4$ along the waveguide, as shown in Fig. 1. It is important to note that, in Eqns. (4-7), the two strong "pumps" at $\omega_{1,2}$ generate an acoustic wave that drives transition between two "signals" at $\omega_{3,4}$. Here the pumps are assumed to have zero detuning but there is a phase mismatch of $\Delta q$ between the signals. Also, the signal fields are assumed to be very weak in comparison with the pumps such that they do not affect the acoustic dynamics generated by the pumps. In previous work [23], it has been shown that the coupling efficiency can be maximized by increasing the ratio of input pump power $P_1(0)$ to $P_2(0)$. This is because the strength of the acoustic field slowly develops and reaches its maximum, instead of decreasing rapidly from $z = 0$ as occurs for balanced pump powers. For the following study, the initial pump powers are fixed at $P_1(0) = 0.999\, P_T$ and $P_2(0) = 0.001\, P_T$ where $P_T$ represents the total input pump power, which unless otherwise stated is taken to be $P_T = 500$ mW. The maximum isolation in dB, $I_{dB}$ is defined as

$$I_{dB} = 10 \log_{10}\left(\frac{P_3(z=L)}{P_3(z=0)}\right) \quad (8)$$

where $L$ is the waveguide length, and $P_3$ is the physical power of the signal at $\omega_3$, with the initial signal power $P_3(0)$ fixed at 1 mW. Equations (4-7) can be solved numerically to obtain $P_i(z)$. It is important to note that while the device insertion loss increases with the optical loss $\alpha_i$, the maximum isolation $I_{dB}$ is independent on $\alpha_i$ because the power flow in both direction experiences the same propagation loss. Hence, $\alpha_i = 0$ is used in our calculation.

## 4.2 Isolation bandwidth and dispersion engineering

The key to expanding the operational bandwidth is maintaining $\Delta q \approx 0$ for a wide range of detuning. The bandwidth $|\delta\omega|$ for which the FIBS can occur is then given by

$$|\delta\omega| = \frac{\pi}{2\, z_{max}} \frac{c}{\Delta n_g} \quad (9)$$

where $c$ is the speed of light in free space and $\Delta n_g$ is the group index difference between the TE$_{11}$ and TE$_{21}$ modes at $\omega_1$. In this equation, $\Delta n_g = n_{g,1} - n_{g,2}$ is the difference in optical group index, where the group index of mode $i$ is given by $n_{g,i} \equiv c/(\frac{d\omega_i}{dk}) = n_i + \omega \frac{dn_i}{d\omega}$, where $n_i(\omega) = ck_i/\omega$ is the effective index of mode $i$. The analytical derivation of $|\delta\omega|$ in Eq. (9) can be found in section 5 of Supplement 1. One way in which the operational bandwidth can be widened is by tailoring the waveguide dispersion (reducing $\Delta n_g$) by changing the waveguide dimensions ($w$ or $t$). The effect of such tuning is shown in Fig. 7: Examining the $|\Delta n_g|$ across a 100 nm wavelength span for three different $w$ (550 nm, 570 nm and 600 nm), we find that waveguides of width $w = 570$ nm yield $|\Delta n_g| < 0.1$ for a 30 nm span around a central wavelength of 1550 nm, with a minimum $|\Delta n_g| \approx 7 \times 10^{-4}$ at $\lambda = 1544$ nm. It is important to note that, when computing the $n_g$, the refractive index of As$_2$Se$_3$ is fixed at 2.84 due to the flat response of material dispersion in the 1500-1600 nm wavelength range (see section 7 of Supplement 1).

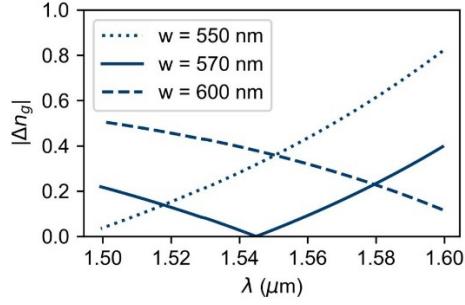

Fig. 7. The group index difference between $TE_{11}$ and $TE_{21}$ mode, $|\Delta n_g|$ versus $\lambda$ at $w = 550$ nm, 570 nm and 600 nm

*4.3 High-performance fabrication-tolerant mode coupler*

The overall performance of the isolator is inseparable from the quality of the mode filters (Fig. 2(a)). Therefore, in addition to optimising the FIBS process in the Brillouin-active region, it is also necessary to have a mode filter with high efficiency and low crosstalk so that the isolation and the operational bandwidth of the whole device are as close as possible to the ideal bandwidth computed from Eqs (4-7). In general, mode (de)multiplexing can be achieved using Y junctions [68], asymmetrical directional couplers (ADCs) [69,70], and multimode interference (MMI) couplers [71]. However, there is a strong motivation to use ADCs because: (1) the methodology for improving the ADCs' dimensional tolerance has been well developed [69,70]; (2) the sharp corners of Y junctions often lead to fabrication challenges; (3) MMI couplers experience higher radiation loss.

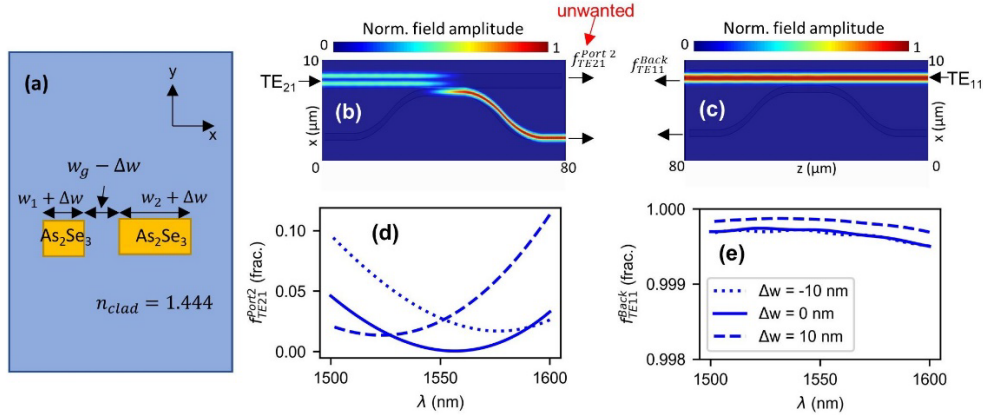

Fig. 8. (a) The image depicts the cross-section of a pair of parallel waveguides with widths $w_1$ and $w_2$ separated by a gap, $w_g$; (b-e) Lumerical 3D FDTD simulation for mode coupler with $w_1 = 562$ nm, $w_2 = 1185$ nm, $w_g = 100$ nm, $L_c = 12$ μm: (b,d) $TE_{21}$ mode injection from FIBS waveguide and output at Port 2 and Port 4 as $TE_{21}$ and $TE_{11}$ modes, with its (b) top view of the field propagation, and (d) the plot of $f_{TE21}^{Port2}$ versus $\lambda$ for different $\Delta w$; (c,e) $TE_{11}$ mode injection from Port 2 and output at FIBS waveguide, with its (c) top view of the field propagation, the spectrum of (e) $f_{TE11}^{Back}$ for $\Delta w = $ -10 nm, 0 nm and 10 nm respectively.

Here, we design a width-tolerant mode coupler for converting between $TE_{11}$ and $TE_{21}$ modes on the $As_2Se_3$ platform. We employ the strategy reported in [69]: ADCs are made of a pair of parallel waveguides with different widths $w_1$ and $w_2$, separated by a gap $w_g$ as shown in Fig. 8(a). The supermode analysis and optimization process for such a ADC based mode coupler is detailed in section 6 of the Supplement 1. The optimized dimension of the mode coupler has already been shown in Fig. 2(b), in which $w_1 = 562$ nm, $w_2 = 1185$ nm, $w_g = 100$ nm. The mode coupler has a coupling length $L_c = 12$ μm.

The simulated transmission spectrum of such an optimized design, subjected to a perturbation $\Delta w$, is plotted in Figs. 8(b-e). From the results plotted in Fig. 8(e), a notably flat transmission response is obtained at the FIBS waveguide, with $f_{TE11}^{Back} \approx 1$, suggesting that the device insertion loss due to mode coupler crosstalk is negligible. However, imperfect mode coupling ($f_{TE21}^{Port2} > 0$) can occur at any wavelength other than the optimum wavelength or when there is a small $\Delta w$ (see Fig. 8(d)). As a consequence, part of the light from TE$_{21}$ mode at $\omega_4$ can output to Port 2 (see Fig. 2(a)), thus reducing the isolation bandwidth estimated from Eq. (4-9).

## 4.4 Computation of overall performance

Based on the mode coupler performance in the previous section, the overall isolation, $I'_{dB}$ as a function of $P_{total}$ and $\lambda$ is then given by

$$I'_{dB}(P_{total}, \lambda) = 10 \log_{10}\left(\frac{P_3(z = L)}{P_3(z = 0)} + f_{TE21}^{Port2}\right) \quad (10)$$

where $\frac{P_3(z=L)}{P_3(z=0)}$ is the linear power fraction of the non-converted TE$_{11}$ mode from the FIBS process that output to Port 2 at $\omega_3$ and $f_{TE21}^{Port2}$ is the linear power fraction of the TE$_{21}$ mode output to Port 2 at $\omega_4$ due to imperfect mode coupling at the mode filter demonstrated in section 4.3.

We further calculate the isolation bandwidth of the 570 nm × 700 nm As$_2$Se$_3$ waveguide using Eq. (10). A two-dimensional parameter sweep of $P_T$ from 0.1 W to 1 W and signal wavelength, $\lambda_s = 2\pi c/\omega_3$ from 1500 nm to 1600 nm was performed for a waveguide length $L = 2$ cm. Such a length is sufficient to achieve complete mode conversion for a reasonable estimate of the experimental $g_0$ ($1 \times 10^3$ m$^{-1}$W$^{-1}$). The resulting values of $I'_{dB}$ as a function of $P_T$ and $\lambda_s$ for $w = 560$ nm, 570 nm, and 580 nm are depicted in Figs. 9(a-c). The results illustrate that while the 30 dB isolation line only exists for high $P_T$, the bandwidths for 10 dB, 20 dB and 30 dB isolation can increase markedly when switching from low to high $P_T$, as a consequence of the reduction in $L$. In addition, at the optimum waveguide width $\Delta w = 0$ and $P_T = 0.5$ W (see Fig. 9(b)), the 10 dB isolation bandwidth exceeds 100 nm while the 30 dB bandwidth is 8 nm (~1000 GHz).

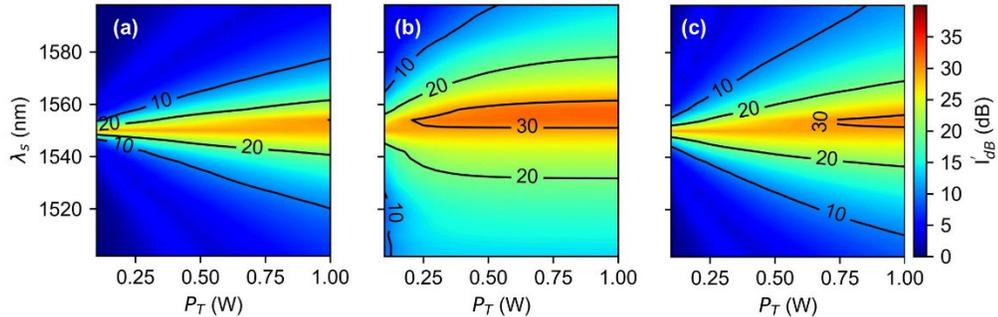

Fig. 9. Overall isolation bandwidth, $I'_{dB}$ as a function of $P_T$ and $\lambda_s$ after incorporating the mode coupler performance at (a) $\Delta w = -10$ nm, (b) $\Delta w = 0$ nm and (c) $\Delta w = 10$ nm.

We can estimate the impact of fabrication uncertainty by incorporating a waveguide width deviation, $\Delta w = \pm 10$ nm into the simulation. As illustrated in Fig. 9(a) and (c), the 30 dB isolation line only exists for high $P_T$, whereas the 10 dB isolation bandwidth at $P_T = 0.5$ W is reduced to 30 nm (~3700 GHz) and 51 nm (~6400 GHz) for $\Delta w = -10$ nm and $\Delta w = +10$ nm respectively.

## 5. Discussion

In principle, the isolation bandwidth could be improved by inserting a number $N$ of coupler replicas before Port 2. The idea is presented in Fig. 10(a). Such a remedial approach leverages the unique nature of the mode coupler that converts the $TE_{21}$ mode in the wider arm to the $TE_{11}$ mode in the narrower arm (see Figs. 8(b, d)) while prohibiting the cross-coupling of the $TE_{11}$ mode from the wider arm (see Figs. 8(c, e)), to maximize the $I'_{dB}$ and $|\delta\omega|$ by modifying the term $f_{TE21}^{Port2}$ into $(f_{TE21}^{Port2})^{N+1}$ in Eq. (10). The investigation of isolation improvement due to coupler addition for specific input conditions ($P_T = 0.5$ W and $\Delta w = 0$ nm) is shown in Fig. 10(b). As predicted, the 30 dB isolation bandwidth increases with $N$ and converges to a maximum bandwidth (the black dashed line) at which the term $f_{TE21}^{Port2}$ approaches zero. In our case, an extra coupler ($N = 1$) is sufficient to retrieve the raw isolation (stated in Eq. 8) from the non-reciprocal mode conversion in the FIBS process. While this approach may be effective, it depends on the efficiency of the mode coupler: using mode couplers with low efficiency would require a large number of coupler replicas, which would significantly increase the device size and overall insertion loss.

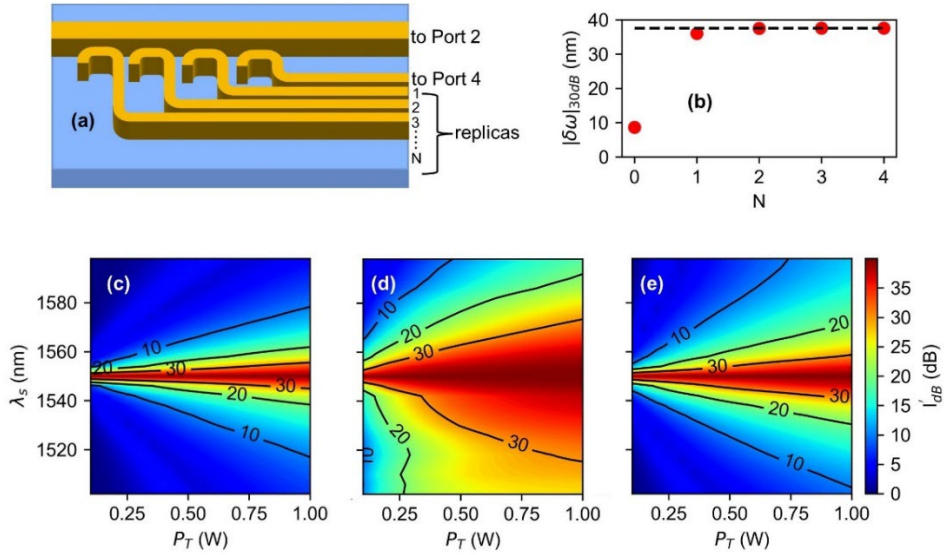

Fig. 10. Improvement of isolation bandwidth: (a) Illustration of the placement of the coupler replicas before Port 2, (b) the plot of 30 dB isolation bandwidth, $|\delta\omega|_{30dB}$ versus the number of coupler replicas, $N$ at $P_{total} = 0.5$ W and $\Delta w = 0$ nm. (c-e) The corrected isolation bandwidth, $I'_{dB}$ as a function of $P_T$ and $\lambda_s$ after incorporating an extra mode coupler ($N = 1$) at (c) $\Delta w = -10$ nm, (d) $\Delta w = 0$ nm and (e) $\Delta w = 10$ nm.

The contour plot of the corrected isolation $I'_{dB}$ as a function of $P_T$ and $\lambda_s$ for $N = 1$ is given in Figs. 10(c-e). At $\Delta w = 0$, the 30 dB isolation bandwidth is increased to 38 nm (~4700 GHz). It can be seen that, in comparison with Figs. 9(a-c), the isolation bandwidth has been considerably improved in which the 30 dB isolation line is now exists for low $P_T$ and even with the presence of $\Delta w = \pm 10$ nm. To illustrate, while the 10 dB isolation bandwidth has been increased to 34 nm (~ 4200 GHz) and 59 nm (~7300 GHz) for $\Delta w = -10$ nm and $\Delta w = +10$ nm respectively, whereas the 30 dB isolation bandwidths are 5 nm (~620 GHz) and 10 nm (~1250 GHz) respectively. These results show that broadband isolation — hundreds of GHz — can be achieved even with the presence of 10 nm structural deviations due to fabrication imperfections.

As a final step in our analysis, we discuss the taper length optimization of the spot size converters and horizontal tapers (see Fig. 2(b)). If these tapers are too short, the abrupt changes in waveguide width can lead to unwanted mode coupling. There are several widths involved in the whole isolator design that must be accommodated: 570 nm in FIBS waveguide, 562 nm and 1185 nm in the mode coupler, and 1900 nm at the waveguide facet for fibre-to-chip coupling. The biggest change of width occurs between Port 3 (or 4) and the narrow arm of the mode coupler (as shown in Fig. 2(b)) where a width expansion from 570 nm to 1900 nm is required. Concerning the width change of 1300 nm, a taper length scan is performed using Lumerical 3D FDTD. It was found that a taper length of 10 μm is safe (> 99.9% transmission) for $TE_{11}$ mode propagation whereas $TE_{21}$ mode will require a taper length of at least 100 μm to suppress any mode coupling (see section 8 of Supplement 1).

It is also worth commenting on the practical viability of the platform proposed in this work. We numerically demonstrate that the narrow (~600 nm wide) selenide ($As_2Se_3$) waveguides have a strong potential for achieving high Brillouin gains ($g_0 \approx 2.4 \times 10^4$ $m^{-1}W^{-1}$ using the phonon viscosity coefficients estimated from 13.2 MHz Brillouin linewidth in Abedin et al. work [54]). From the literature, the $As_2Se_3$ family (GeAsSe [72] or Ag-$As_2Se_3$ [73]) possesses a higher intrinsic material absorption at 1550 nm than $As_2S_3$ [74]. Besides, the waveguide sidewall roughness inherited from the dry-etching can also lead to large propagation loss and mode coupling, in particular for the higher-order optical modes. The propagation loss can range from sub dB/cm to a few dB/cm depending on the etching recipe and waveguide geometry [74]. Such losses are identical in both propagation directions and the overall isolation will not be affected when $\Delta q_{pm} = 0$. However, they can lead to a shorter effective length, from which a higher pump power will be required to maintain such a large device operational bandwidth provided the damage threshold of the waveguide is not exceeded. Therefore, engineering the waveguide loss should be carried out in the future to improve energy efficiency of the proposed Brillouin isolators.

## 6. Conclusion

To summarise, we have carried out an extensive study on an SBS-based optical isolator from both dimensional optimization and fabrication error perspectives. We have shown that the acoustic confinement from FIBS process can be improved by engineering the waveguide dimension and utilising an appropriate optical mode transition. More interestingly, the polarization of the acoustic displacement field relies on the choice of the spatial optical mode pairs. Through waveguide dispersion tailoring, a 30-dB isolation bandwidth of 38 nm (~4700 GHz) can be achieved in a 570 nm × 700 nm $As_2Se_3$ waveguide with a pump power of 500 mW while the bandwidth is reduced to 5 – 10 nm at a width error of ±10 nm. Together with fabrication tolerant mode couplers and the coupler insertion scheme proposed, we have shown that the overall isolation achieved can be solely dependent on the raw isolation from the FIBS process. The simulation and modelling efforts in this work take us one step closer to realizing the on-chip Brillouin-based optical isolator.

**Funding.** This research was supported by the Australian Government through the Australian Research Council's Discovery Projects funding scheme (project DP200101893).

**Disclosures.** The authors declare no conflicts of interest.

**Supplemental document.** See Supplement 1 for supporting content.

# Optimizing performance for on-chip SBS-based isolator: supplemental document

Non-reciprocal optical components such as isolators and circulators are crucial for preventing catastrophic back-reflection and controlling optical crosstalk in photonic systems. While non-reciprocal devices based on Brillouin intermodal transitions have been experimentally demonstrated in chip-scale platforms, harnessing such interactions has required a suspended waveguide structure, which is challenging to fabricate and is potentially less robust than a non-suspended structure, thereby limiting the design flexibility. In this paper, we numerically investigate the feasibility of a Brillouin-based isolation scheme in which a dual-pump-driven optoacoustic interaction is used to excite confined acoustic waves in a traditional ridge waveguide. We find that acoustic confinement, and therefore the amount of Brillouin-driven mode conversion, can be enhanced by selecting an appropriate optical mode pair and waveguide geometry of two binary arsenic based chalcogenide platforms. Further, we optimize the isolator design in its entirety, including the input couplers, mode filters, the Brillouin-active waveguide as well as the device fabrication tolerances. We predict such a device can achieve 30 dB isolation over a 38 nm bandwidth when 500 mW pump power is used; in the presence of a ± 10 nm fabrication-induced width error, such isolation can be maintained over a 5-10 nm bandwidth.

## 1. INTRODUCTION

This document provides supplementary information for the paper "Optimizing performance for on-chip SBS based isolator". Section 2 details the simulation setup of the optical and acoustic mode solver in the Numerical Brillouin Analysis Tool (NumBAT). The forward intermodal Brillouin scattering (FIBS) process in this work involves two pump and two signal optical waves and thus the degree of non-reciprocal mode conversion can be obtained by solving four-mode-coupling equations. The derivation of such equations is given in section 3 and 4. Then section 5 shows the analytical derivation of the isolation bandwidth that is useful for estimating the operational bandwidth of the proposed isolator. The design and optimization process of a width-tolerant mode (de-)multiplexers in Arsenic triselenide platform is detailed in section 6 of this document.

## 2. NUMBAT - THE NUMERICAL BRILLOUIN ANALYSIS TOOL

NumBAT is an open source freeware available on https://github.com/michaeljsteel/NumBAT. It integrates electromagnetic and acoustic mode solvers to calculate the interactions of optical and acoustic waves in waveguides. This section provides the additional information with regards to the setup of the NumBAT simulation and mesh.

In general, the simulation algorithm of NumBAT for this work consists of three steps. First, all supported optical modes in an arbitrary multimode waveguide at a specific free space optical wavelength (1.55 µm) are obtained using the Finite Element Method (FEM). The normalized electric field distribution of the fundamental and higher order quasi-transverse electric (TE) modes is illustrated in Fig. 3 of the paper. The quasi-transverse magnetic (TM) modes have very similar profiles as their TE counterparts but with a dominant electric field component in the y-direction. Second, by specifying an optical mode pair of interest, the acoustic wavenumber, $q$ can be determined to solve for the acoustic modes. At a particular $q$, a number of acoustic modes at different frequencies (Brillouin shift $\Omega$) can be found. The strength of the Brillouin coupling for each of these modes is then computed through standard expressions based on optoacoustic mode overlap integrals [1, 2].

Both simulation domain and mesh size play an important role in the FEM analysis. Large domain size is necessary to avoid any back-reflection from the boundary of the simulation region, in particular the less-confined transverse acoustic modes from the FIBS process. The required domain size can be well approximated using the transverse decay constant, $\kappa$ and the decay length, $L_{decay}$ of such modes:

$$\kappa = \frac{\Omega}{v_{s,cl}} \sqrt{\left(\frac{v_{s,cl}}{v_p}\right)^2 - 1} \tag{S1}$$

$$L_{decay} = \frac{1}{\kappa} \tag{S2}$$

For all modes in Fig. 4(e-h) of the paper, we take $\Omega = 1.3$ GHz as representative of the range of values from 0.9 to 1.7 GHz, the shear speed $v_{s,cl} \approx 3766$ ms$^{-1}$ for silica cladding, and thus for a reasonably confined mode ($v_p < 3400$ ms$^{-1}$), a domain size of 15.5 µm (ten times the free-space optical wavelength) is sufficient to cover the $L_{decay}$ of the displacement field. In terms of mesh size, the boundary of the simulation region takes a maximum cell size of 800 nm, and it is gradually decreased to 0.5 % of that at the waveguide boundaries, with a minimum cell size of 3 - 15 nm depending on the waveguide geometry.

## 3. NORMALISATION OF THE COUPLED MODE EQUATIONS

Here we write down the coupled mode equations for the on-chip isolator. Overall we start from the coupled mode equations provided in Wolff et al. work [1], and use the normalisation in Nieve's thesis [3], followed by adding an additional two optical fields that are being driven by the acoustic mode, as in the Poulton et al. work [4]. Consider a waveguide oriented along the z axis, with optical fields propagating in the positive z direction. The (unitless) coupled mode equations for the envelope fields are (Equations (71-73) in reference [1]):

$$\frac{\delta A_1}{\delta z} + \frac{1}{v_1}\frac{\delta A_1}{\delta t} + \frac{\alpha_1}{2}A_1 = -i\omega_1 \frac{\tilde{Q}_1}{P_1} B A_2 \tag{S3}$$

$$\frac{\delta A_2}{\delta z} + \frac{1}{v_2}\frac{\delta A_2}{\delta t} + \frac{\alpha_2}{2}A_2 = -i\omega_2 \frac{\tilde{Q}_2}{P_2} B^* A_1 \tag{S4}$$

$$\frac{\delta B}{\delta z} + \frac{1}{v_a}\frac{\delta B}{\delta t} + \frac{\alpha_a}{2}B = -i\Omega \frac{\tilde{Q}_a}{P_a} A_1^* A_2 \tag{S5}$$

Here $A_1$ describes the envelope of the higher frequency mode (the pump), $A_2$ describes the lower frequency envelope (the Stokes), and $B$ describes the envelope of the acoustic field. $v_{1,2,a}$ are the group velocities of the modes, $\alpha_{1,2,a}$ are the modal power losses in units of m$^{-1}$, and the coupling constants $\tilde{Q}_{1,2,a}$ are as defined as in the reference [1] (equations (29,30,46)). $P_{1,2,a}$ are normalisation constants which set the powers of the modes such that the power in mode 1 at point $z$ and time $t$ is equal to $P_1|A_1(z,t)|^2$ watts.

We now define new envelopes with squared amplitudes representing physical powers:

$$a_1(z,t) = \sqrt{|P_1|} A_1(z,t), \ a_2(z,t) = \sqrt{|P_2|} A_2(z,t), \ b(z,t) = \sqrt{|P_a|} B(z,t) \quad (S6)$$

Substituting into (S1)-(S3) we get

$$\frac{\delta a_1}{\delta z} + \frac{1}{v_1}\frac{\delta a_1}{\delta t} + \frac{\alpha_1}{2} a_1 = -i\omega_1 \frac{\tilde{Q}_1}{\sqrt{|P_1 P_2 P_a|}} b a_2 \quad (S7)$$

$$\frac{\delta a_2}{\delta z} + \frac{1}{v_2}\frac{\delta a_2}{\delta t} + \frac{\alpha_2}{2} a_2 = -i\omega_2 \frac{\tilde{Q}_2}{\sqrt{|P_1 P_2 P_a|}} b^* a_1 \quad (S8)$$

$$\frac{\delta b}{\delta z} + \frac{1}{v_a}\frac{\delta b}{\delta t} + \frac{\alpha_a}{2} b = -i\Omega \frac{\tilde{Q}_a}{\sqrt{|P_1 P_2 P_a|}} a_1^* a_2 \quad (S9)$$

or, using $\tilde{Q}_1 = \tilde{Q}_2 = \tilde{Q}_a$ (which comes from (68) in the reference [1], and reflects the absence of irreversible processes) we find

$$\frac{\delta a_1}{\delta z} + \frac{1}{v_1}\frac{\delta a_1}{\delta t} + \frac{\alpha_1}{2} a_1 = -i\omega_1 Q^* b a_2 \quad (S10)$$

$$\frac{\delta a_2}{\delta z} + \frac{1}{v_2}\frac{\delta a_2}{\delta t} + \frac{\alpha_2}{2} a_2 = -i\omega_2 Q b^* a_1 \quad (S11)$$

$$\frac{\delta b}{\delta z} + \frac{1}{v_a}\frac{\delta b}{\delta t} + \frac{\alpha_a}{2} b = -i\Omega Q a_1^* a_2 \quad (S12)$$

where

$$Q = \frac{\tilde{Q}_1}{\sqrt{|P_1 P_2 P_a|}} \quad (S13)$$

Here the physical power in W carried in the optical fields is $|a_{1,2}(z,t)|^2$, and the power in the acoustic field is $|b(z,t)|^2$.

The value $\tilde{Q}_1$ is given by NumBAT, which also outputs the mode normalisation powers $P_{1,2,a}$. From equation (2.60) in reference [3], we could also calculate $Q$ from

$$Q = \left[ 2v_1 v_2 v_a \rho_0 \Omega^2 \left( A_{eff}^{ao} \right)^2 \right]^{-1}$$

where $\rho_0$ is the core density and $A_{eff}^{ao}$ is the acousto-optic effective area.

In the steady state and with zero detuning the coupled mode equations become

$$\frac{\delta a_1}{\delta z} + \frac{\alpha_1}{2} a_1 = -i\omega_1 Q^* b a_2 \quad (S14)$$

$$\frac{\delta a_2}{\delta z} + \frac{\alpha_2}{2} a_2 = -i\omega_2 Q b^* a_1 \quad (S15)$$

$$\frac{\delta b}{\delta z} + \frac{\alpha_a}{2} b = -i\Omega Q a_1^* a_2 \quad (S16)$$

Equation (S16) can be solved analytically in terms of the optical fields (this is most easily done just using an integrating factor):

$$b(z) = -i\Omega Q \int_0^\infty dz' a_1^*(z - z') a_2(z - z') \exp\left[-\frac{\alpha_a}{2} z'\right] \quad (S17)$$

We can go a bit further (as in the original isolator paper) if we assume that all the optical fields and their derivatives vary slowly in comparison with the acoustic lifetime. The acoustic field then "has time to follow" the beat pattern caused by interference between the optical fields. To do this we use a Taylor series of $f(z - z_0)$ about $z$ in (S17) and so end up with

$$b(z) \approx -i\frac{2\Omega Q}{\alpha_a} a_1^*(z) a_2(z) \quad (S18)$$

Substituting this into equations (S14) and (S15), and assuming $\omega := \omega_1 \approx \omega_2$ we obtain

$$\frac{\delta a_1}{\delta z} + \frac{\alpha_1}{2} a_1 = -\frac{g_0}{2}|a_2|^2 a_1 \tag{S19}$$

$$\frac{\delta a_2}{\delta z} + \frac{\alpha_2}{2} a_2 = -\frac{g_0}{2}|a_1|^2 a_2 \tag{S20}$$

where

$$g_0 = \frac{4\omega\Omega|Q|^2}{\alpha_a} \tag{S21}$$

is the Brillouin gain in m$^{-1}$W$^{-1}$. This is the same expression as in (91) of the reference [1], noting that the paper defines the acoustic loss $\alpha$ as an envelope amplitude loss rather than a power amplitude loss, thereby leading to a factor 2 in the numerator of (S21) rather than the factor 4 that we have here. The quantity $g_0$ should be the number provided by NUMBAT.

## 4. FOUR-MODE COUPLING EQUATIONS

We now consider the case where we have four modes coupled together (see Fig. S4), with two strong "pumps" at $\omega_{1,2}$ which generate an acoustic wave that drives transitions between two "signals" at $\omega_{3,4}$. If we assume that there is zero detuning of the pumps but that there is a phase mismatch of $\Delta q$ between the signals, then the system of coupled equations is

$$\frac{\delta a_1}{\delta z} + \frac{\alpha_1}{2} a_1 = -i\omega_1 Q^* b a_2 \tag{S22}$$

$$\frac{\delta a_2}{\delta z} + \frac{\alpha_2}{2} a_2 = -i\omega_2 Q b^* a_1 \tag{S23}$$

$$\frac{\delta b}{\delta z} + \frac{\alpha_a}{2} b = -i\Omega Q a_1^* a_2 \tag{S24}$$

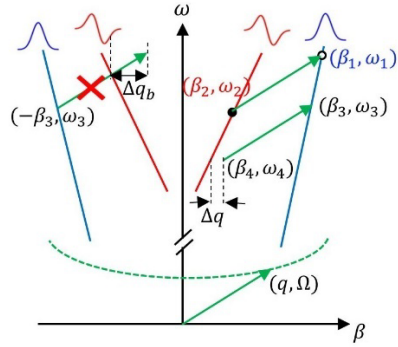

Fig. S1. Dispersion diagram of intermodal forward Brillouin scattering

for the pumps and acoustic mode. The equations for the signals are basically the same as for the optical modes (S22) and (S23) given that $\omega_3$ is the higher frequency and $\omega_4$ is the lower frequency. The main differences are that there is a phase-mismatch $\Delta q$ that comes from the fact that the dispersion curves are not necessarily parallel. The equations are

$$\frac{\delta a_3}{\delta z} + \frac{\alpha_3}{2} a_3 = -i\omega_3 Q^* a_4 b \exp(-i\Delta q z) \tag{S25}$$

$$\frac{\delta a_4}{\delta z} + \frac{\alpha_4}{2} a_4 = -i\omega_4 Q^* a_3 b \exp(-i\Delta q z) \tag{S26}$$

We now make the same slowly-varying pulse assumption as in the previous section, assume that all the optical frequencies are sufficiently close together that we can write $\omega_3 \approx \omega_4 \approx \omega$, and note that the signal fields are very weak in comparison with the pumps such that they do not affect the acoustic dynamics generated by the pumps. We can then write the equations for the signal fields as

$$\frac{\delta a_3}{\delta z} + \frac{\alpha_3}{2} a_3 = -i\omega_3 Q^* \left(-i\frac{2\Omega Q}{\alpha_a} a_1^* a_2\right) \exp(-i\Delta q z) a_4$$

$$= -\frac{g_0}{2} a_1^* a_2 \exp(-i\Delta q z) a_4 \tag{S27}$$

$$\frac{\delta a_4}{\delta z} + \frac{\alpha_4}{2} a_4 = -i\omega_4 Q^* \left(i\frac{2\Omega Q^*}{\alpha_a} a_1 a_2^*\right) \exp(-i\Delta q z) a_3$$

$$= \frac{g_0}{2} a_1 a_2^* \exp(-i\Delta q z) a_3 \tag{S28}$$

The system of equations that we have to solve is then

$$\frac{\delta a_1}{\delta z} + \frac{\alpha_1}{2} a_1 = -\frac{g_0}{2} |a_2|^2 a_1 \tag{S29}$$

$$\frac{\delta a_2}{\delta z} + \frac{\alpha_2}{2} a_2 = \frac{g_0}{2} |a_1|^2 a_2 \tag{S30}$$

$$\frac{\delta a_3}{\delta z} + \frac{\alpha_3}{2} a_3 = -\frac{g_0}{2} a_1^* a_2 \exp(-i\Delta q z) a_4 \tag{S31}$$

$$\frac{\delta a_4}{\delta z} + \frac{\alpha_4}{2} a_4 = \frac{g_0}{2} a_1 a_2^* \exp(-i\Delta q z) a_3 \tag{S32}$$

with the subsidiary equation

$$b(z) \approx -i\frac{2\Omega Q}{\alpha_a} a_1^*(z) a_2(z) \tag{S33}$$

for the acoustic field.

## 5. ISOLATION BANDWIDTH

The isolation process is based on efficient coupling between the signal fields mediated by an acoustic field generated by the pump fields with wavenumber $q_p = |\beta_1 - \beta_2|$. The signals beat at $q_s = |\beta_3 - \beta_4|$ and so for effective cross coupling we need that $\exp(i|q_p - q_s|L) \ll \pi/2$ or

$$\Delta q = |q_p - q_s| \lesssim \frac{\pi}{2L}. \tag{S34}$$

We'll call the mode with pump 1 and signal 1 mode A, and the other one mode B.

Given pump 2 is a Brillouin shift away from $\omega_1$ at $\omega_2 = \omega_1 + \Omega$, signal 3 is at an unknown frequency offset $\omega_3 = \omega_2 + \delta\omega$ and signal 4 is at $\omega_4 = \omega_3 + \Omega$. Therefore, $\Delta q$ can be expressed in terms of $\omega_1$ only. Now we just Taylor expand everything and look for a constraint on $|\delta\omega|$, using the standard notation $\beta_n(\omega_0) = \frac{d^n\beta(\omega)}{d\omega^n}|_{\omega=\omega_0}$

$$\Delta q$$
$$= q_s - q_p$$
$$= |\beta_3 - \beta_4| - |\beta_1 - \beta_2|$$
$$= |\beta^A(\omega_3) - \beta^B(\omega_4)| - |\beta^A(\omega_1) - \beta^B(\omega_2)|$$
$$= |\beta^A(\omega_1 + \delta\omega) - \beta^B(\omega_1 + \delta\omega + \Omega)| - |\beta^A(\omega_1) - \beta^B(\omega_1 + \Omega)| \tag{S35}$$

$$\approx \left| \begin{pmatrix} \beta^A(\omega_1) + \delta\omega\beta_1^A(\omega_1) \\ + \frac{\delta\omega^2}{2}\beta_2^A(\omega_1) \end{pmatrix} - \begin{pmatrix} \beta^B(\omega_1) + (\delta\omega + \Omega)\beta_1^B(\omega_1) + \\ \frac{(\delta\omega+\Omega)^2}{2}\beta_2^B(\omega_1) \end{pmatrix} \right| - \left| \beta^A(\omega_1) - \left( \beta^B(\omega_1) + \Omega\beta_1^B(\omega_1) + \frac{\Omega^2}{2}\beta_2^B(\omega_1) \right) \right|$$
(S36)

$$= \left| \left(\beta^A(\omega_1) - \beta^B(\omega_1)\right) + \delta\omega\left(\beta_1^A(\omega_1) - \beta_1^B(\omega_1)\right) - \Omega\beta_1^B(\omega_1) + \left(\frac{\delta\omega^2}{2}\beta_2^A(\omega_1) - \frac{(\delta\omega-\Omega)^2}{2}\beta_2^B(\omega_1)\right) \right| - \left| \beta^A(\omega_1) - \left(\beta^B(\omega_1) + \Omega\beta_1^B(\omega_1) + \frac{\Omega^2}{2}\beta_2^B(\omega_1)\right) \right|$$
(S37)

Dropping the quadratic terms and noting that the two terms in the second line are both positive

$$\Delta q \approx \delta\omega\left(\beta_1^A(\omega_1) - \beta_1^B(\omega_1)\right) - \Omega\beta_1^B(\omega_1) + \Omega\beta_1^B(\omega_1)$$

$$= \delta\omega\left(\beta_1^A(\omega_1) - \beta_1^B(\omega_1)\right) \tag{S38}$$

and so the bandwidth condition is simply

$$|\delta\omega| \lesssim \frac{\pi}{2L} \frac{1}{\left|\beta_1^A(\omega_1) - \beta_1^B(\omega_1)\right|}$$
(S39)

$$= \frac{\pi}{2L} \frac{c}{\left|n_g^A(\omega_1) - n_g^B(\omega_1)\right|}$$
(S40)

## 6. DESIGN AND OPTIMIZATION OF WIDTH-TOLERANT MODE COUPLER

This section provides detailed optimization process and dimensional analysis of the mode splitter based on asymmetrical directional couplers (ADCs) that will be used in our study. ADCs are made of a pair of parallel waveguides with different widths $w_1$ and $w_2$, separated by a gap $w_g$ as shown in Fig. S2(a). When the modal effective index matching between the two modes is attained, the TE$_{11}$ mode in the narrower arm and the TE$_{21}$ mode in the wider arm will be efficiently coupled. The mode coupling efficiency is given by

$$K = A\sin^2\frac{\pi\Delta n}{\lambda}L = A\sin^2\left(\frac{\pi}{2}\frac{L}{L_c}\right)$$
(S41)
.

where $\Delta n$ is the effective index difference between the odd and even supermodes, $L_c$ is the beat length required to cross couple the power completely to the adjacent waveguide and $L$ is the length of the parallel waveguides. The term $A$ in S41 represents a mode overlap integral which can be formulated as [5]

$$A = \left| Re\left[ \frac{\int(E_x^{\pm} H_y^m dS) \int(E_x^m H_y^{\pm} dS)}{\int(E_x^{\pm} H_y^{\pm} dS) \int(E_x^m H_y^m dS)} \right] \right|$$

(S42)

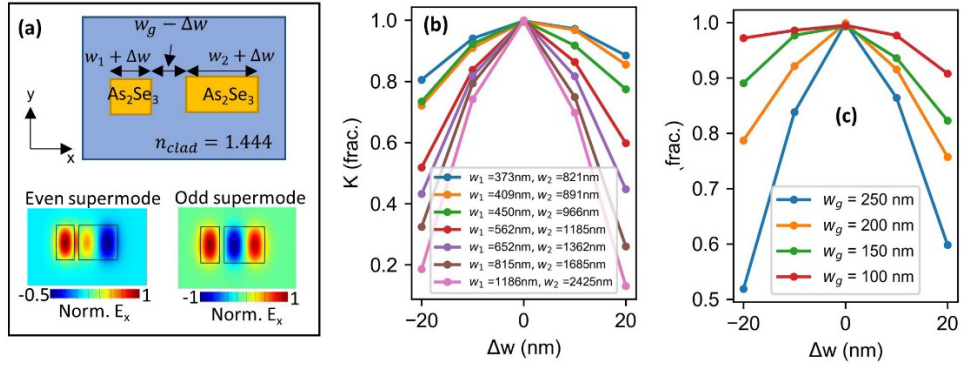

Fig. S2. Determination of mode coupling efficiency, $K$ with the presence of width errors, $\Delta w$ via supermode analysis: (a) the top image depicts the cross-section of a pair of parallel waveguides with widths $w_1$ and $w_2$ separated by a gap, $w_g$ while the two bottom images show the typical profile of the x-component of the electric field, Ex for even and odd supermodes; (b) plot of $K$ versus $\Delta w$ for different $w_1 - w_2$ combinations at $w_g$ = 250 nm; (c) plot of $K$ versus $\Delta w$ for different $w_g$ when $w_1$ = 562 nm and $w_2$ = 1185 nm.

Where $\int dS$ represents the area integral over the waveguide cross-section, Ex is the x-component of the electric field and Hy is the y-component of the magnetic field. The subscript ± indicates the modes derived from the sum or difference between the odd and even supermodes as shown in Fig. S2(a) while subscript m represents their respective isolated waveguide modes. To determine all parameters listed in S41 and S42, a supermode analysis was performed using the Lumerical Finite Difference Eigenmode (FDE) solver.

From S41 it can be seen that $A$ and $Lc$ are the key parameters affecting $K$. At nominal widths $w_1$ and $w_2$, $A$=1, $L_c$=$L$, and thus $K$=1. If a width error $\Delta w$ is introduced, the width of the two waveguides expands in both directions and become $w_1$+ $\Delta w$ and $w_1$+ $\Delta w$, and the gap shrinks to $w_g$- $\Delta w$, causing $A$ < 1 and $L_c \neq L$. The central aim is therefore to come up with a design that is insensitive to $\Delta w$: specifically, the optimal design should maintain $A$ = 1 and $L_c$=$L$ in the presence of a small $\Delta w$. To assess if there exists such a tolerant design, we fix $w_g$ = 250 nm and determine $K$ at -20 nm < $\Delta w$ < 20 nm for seven width combinations satisfying the phase-matching condition, i.e., each of these width pairs has a $TE_{11}$ mode at $w_1$ and $TE_{21}$ mode at $w_1$ that share the same normalized optical wavevector $k/2\pi$. A general trend is obtained in Fig. S2(b) that the dimensional tolerance can be improved by selecting a narrower width pair, but the trade-off is that they are more prone to experiencing a high sidewall scattering loss. Hence, ($w_1$ = 562 nm, $w_2$ = 1185 nm) design is selected for the following study. Now we examine the effect of $w_g$ on the mode coupler performance. In the simulation, $w_g$ of 250 nm, 200 nm, 150 nm, and 100 nm are used, which corresponds to L of 68 μm, 45 μm, 29 μm and 18 μm. The plot in Fig. S2(c) suggests that a smaller $w_g$ can lead to a better dimensional tolerance. Despite that, $w_g$ < 100 nm is not recommended to ensure that L is nonzero.

While the supermode analysis described above allows the prediction of the dimensional tolerance in a pair of parallel waveguides, in a realistic device S-bends should be added to separate the two parallel waveguides for preventing further mode coupling when the maximum cross coupling is attained. To compute the effect that these S-bends might have, a three-dimensional finite-difference time-domain simulation of the full device is carried out using Lumerical 3D FDTD. To begin with, $TE_{21}$ mode was injected from the FIBS waveguide, then the transmission spectrum of $TE_{21}$ mode from Port 2, $f_{TE21}^{Port2}$ and TE11 mode from Port 4, $f_{TE11}^{Port4}$ were recorded. The top view of the field propagation in such a simulation is depicted in Fig. S3(a). We found that the addition of an S-bend with a fixed radius of 50 μm and a final angle of 20⁰ reduces the priorly estimated $L$ to 64 μm, 40 μm, 23 μm and 12 μm for $w_g$ = 250 nm, 200 nm, 150 nm, and 100 nm respectively. The plot of coupling efficiency $f_{TE11}^{Port4}$ versus $\Delta w$ for different $w_g$ in Fig. S2(b) shows a similar trend to that of Fig. S2(c), confirming that a reduction of $w_g$ can help realize a width-insensitive design.

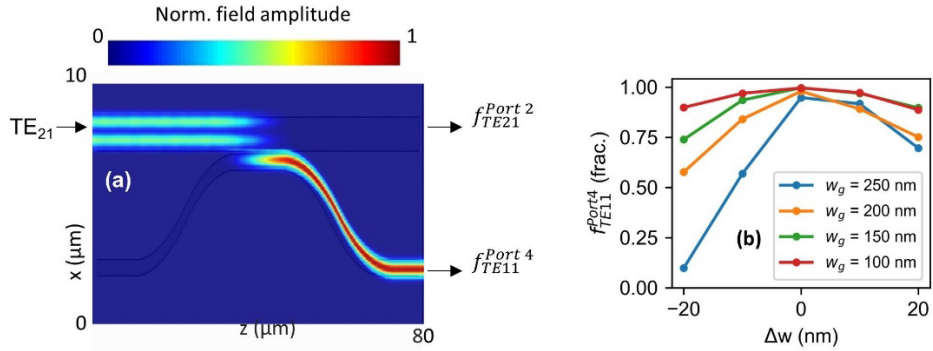

Fig. S3. Lumerical 3D FDTD simulation for mode coupler with $w_1$ = 562 nm and $w_2$ = 1185 nm when $TE_{21}$ mode is injected from FIBS waveguide and output at Port 2 and Port 4 as $TE_{21}$ and $TE_{11}$ modes, with its (a) top view of the field propagation and (b) the plot of $f_{TE11}^{Port4}$ versus $\Delta w$ for different $w_g$.

## 7. MATERIAL DISPERSION OF $AS_2S_3$ AND $AS_2SE_3$

Apart from waveguide and mode dispersion, the material dispersion is also playing an important role in the difference of group index between the two optical modes of interest $\Delta n_g$ and thus the isolation bandwidth. Figure S4 illustrates the dispersion relation of the two arsenic based platform used in this study. The refractive index measurement is carried out using reflectometry technique. The results show that the refractive index of these two films in the wavelength range (1500 nm - 1600 nm) has a reasonably flat response, with no changes observed in the second decimal places of the refractive index. Therefore the effect of material dispersion is negligible as compared with the waveguide and mode dispersion. In our isolation bandwidth study, the material refractive index of $As_2Se_3$ is set at 2.84 for the entire wavelength range of study (1500 nm -1600 nm).

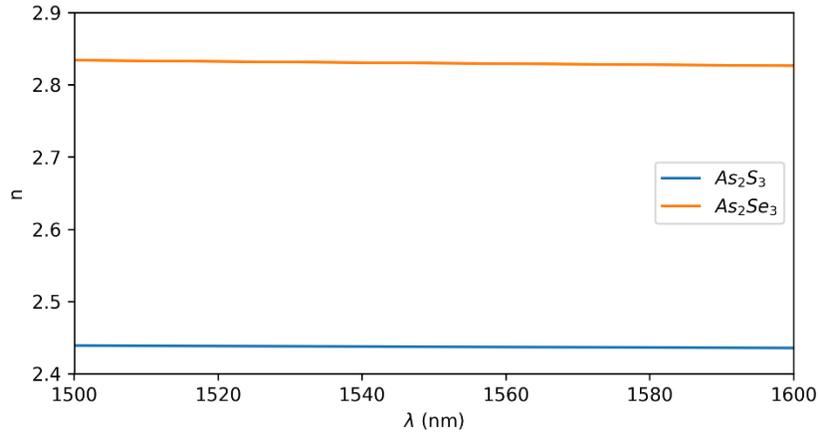

Fig. S4. Measured dispersion relation of As2S3 and As2Se3 at 1500-1600 nm wavelength range

## 8. DESIGN AND OPTIMIZATION OF WAVEGUIDE TAPER

Previous simulation shows that the Brillouin active waveguide (NMRC) and the linear mode couplers (MC1 and MC2) mentioned in the paper are of different waveguide width. Furthermore, the waveguide width at the chip end facet has to be increased to mode matched the butt coupled fibers. Therefore a horizontal waveguide taper is required to connect different components on a single chip.

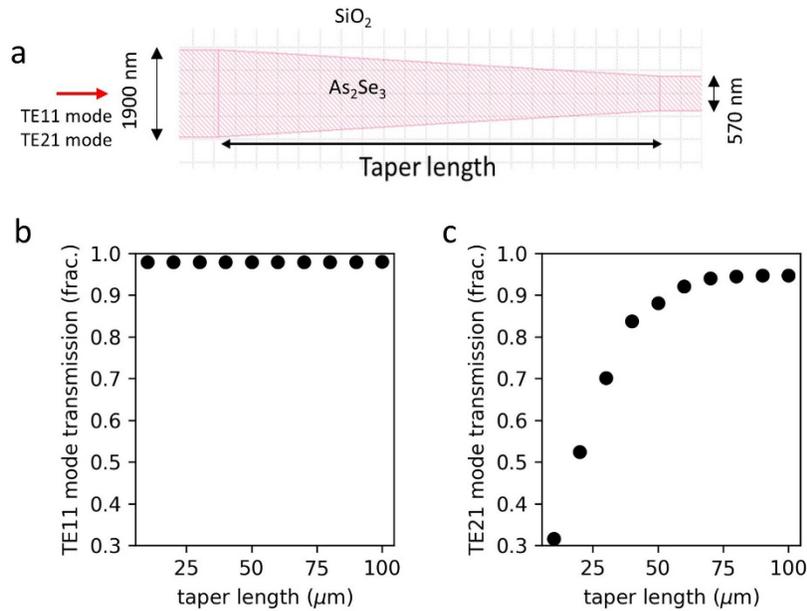

Fig. S5. Lumerical 3D FDTD simulation for $As_2Se_3$ waveguide taper: (a) the taper structure used in the simulation, (b) plot of $TE_{11}$ mode transmission versus taper length and (c) plot of $TE_{21}$ mode transmission versus taper length

This section discusses the waveguide taper design used in the proposed Brillouin isolator. In principle, unsuitable rate of change of width along the taper can cause significant transition loss owing to radiation loss, reflection or mode coupling. Therefore, the taper length for different optical spatial modes must be carefully studied to maximize the taper efficiency. Here we run the Lumerical 3D FDTD simulation to confirm the taper length required for mode transition from a 1900 nm wide to a 570 nm wide $As_2Se_3$ waveguide as shown in Fig. S5(a). By launching two different spatial modes $TE_{11}$ and $TE_{21}$ at the taper input, the transmission corresponding to each spatial modes at the output are plotted in (b) and (c). The results clearly show that the $TE_{11}$ mode records a transmission over 99.8% for the entire taper length range whereas $TE_{21}$ mode transmission increases with the taper length and eventually saturates near 80 µm.